\documentclass[a4paper,twocolumn,10pt,accepted=2020-09-13]{quantumarticle}
\pdfoutput=1
\usepackage[utf8]{inputenc} 
\usepackage[english]{babel} 
\usepackage[T1]{fontenc}
\usepackage{hyperref}
\usepackage{graphicx}  
\usepackage[numbers,sort&compress]{natbib}
\usepackage{subfigure}
\usepackage{amssymb}   
\usepackage{amsmath}
\usepackage{mathtools}
\usepackage{color}
\usepackage[normalem]{ulem}

\begin{document}

\title{Accelerated variational algorithms for digital quantum simulation of the many-body ground states}

\author{Chufan Lyu}
\email{chufan.lyu@std.uestc.edu.cn}
\orcid{0000-0002-8708-9073} 
\affiliation{Institute of Fundamental and Frontier Sciences, University of Electronic Science and Technology of China, Chengdu 610051, China}

\author{Victor Montenegro}
\email{vmontenegro@uestc.edu.cn}
\orcid{0000-0003-3846-3863} 
\affiliation{Institute of Fundamental and Frontier Sciences, University of Electronic Science and Technology of China, Chengdu 610051, China}

\author{Abolfazl Bayat}
\email{abolfazl.bayat@uestc.edu.cn}
\orcid{0000-0003-3852-4558} 
\affiliation{Institute of Fundamental and Frontier Sciences, University of Electronic Science and Technology of China, Chengdu 610051, China}

\begin{abstract}
One of the key applications for the emerging quantum simulators is to emulate the ground state of many-body systems, as it is of great interest in various fields from condensed matter physics to material science. Traditionally, in an analog sense, adiabatic evolution has been proposed to slowly evolve a simple Hamiltonian, initialized in its ground state, to the Hamiltonian of interest such that the final state becomes the desired ground state. Recently, variational methods have also been proposed and realized in quantum simulators for emulating the ground state of many-body systems. Here, we first provide a quantitative comparison between the adiabatic and variational methods with respect to required quantum resources on digital quantum simulators, namely the depth of the circuit and the number of two-qubit quantum gates. Our results show that the variational methods are less demanding with respect to these resources. However, they need to be hybridized with a classical optimization which can converge slowly. Therefore, as the second result of the paper, we provide two different approaches for speeding the convergence of the classical optimizer by taking a good initial guess for the parameters of the variational circuit. We show that these approaches are applicable to a wide range of Hamiltonian and provide significant improvement in the optimization procedure. 
\end{abstract}

\maketitle

\section{Introduction}\label{section:Introduction}
Simulating strongly correlated many-body systems at and out of equilibrium is one of the key tasks in condensed matter physics. On classical computers, simulating a general quantum many-body system is inherently intractable, due to the exponential growth of the Hilbert space with respect to the system size. The only genuine approach for understanding  a quantum system is to emulate its behavior on a quantum simulator, which is a quantum device with more controllability and versatility~\cite{lloyd1996universal,buluta2009quantum,cirac2012goals,georgescu2014quantum}. Quantum simulators are now rapidly emerging in various physical systems, including optical lattices~\cite{gross2017quantum,schreiber2015observation,bordia2017probing,fukuhara2013quantum,fukuhara2013microscopic}, Rydberg atoms~\cite{bernien2017probing}, ion traps~\cite{lanyon2011universal,zhang2017observation,hempel2018quantum,kokail2019self,shen2017quantum}, photonic circuits~\cite{aspuru2012photonic,wang2019integrated}, quantum dot arrays~\cite{hensgens2017quantum}, dopants in silicon~\cite{salfi2016quantum,salfi2016quantum2} and superconducting devices~\cite{hempel2018quantum,colless2018computation,arute2020hartree,ye2019propagation,xu2018emulating,yan2019strongly,las2014digital,salathe2015digital,barends2016digitized,roushan2017spectroscopic,o2016scalable,kandala2017hardware}. There are two different approaches for implementing quantum simulation, namely analog and digital. In the analog approach, the particles interact via some Hamiltonian, possibly time-dependent, and the quantum simulation is the continuous evolution of this system~\cite{bordia2017probing,fukuhara2013quantum,fukuhara2013microscopic,roushan2017spectroscopic,xu2018emulating,yan2019strongly}. Although, reasonably strong noise can be tolerated ~\cite{petrosyan2010state,yang2010spin,farooq2015adiabatic}, the analog quantum simulation is not universal, in the sense that only limited types of unitary operations can be implemented. In contrast, the digital quantum simulation is implemented through quantum gates acting on either one or two qubits in the system~\cite{las2014digital,salathe2015digital,barends2016digitized}. The main advantage of digital quantum simulation is the capability of performing universal computation, i.e. implementing arbitrary unitary operations~\cite{nielsen2002quantum}. The major drawback is the high fragility of such simulators to noise and their demand for complex error correction schemes~\cite{suter2016colloquium}. 

Many important phenomena in condensed matter physics, material science, and chemistry are explained from the ground state of a certain Hamiltonian. This includes electronic structures of matter~\cite{helgaker2014molecular}, molecular formations~\cite{jensen2017introduction}, magnetization~\cite{craik2003magnetism} and quantum phase transitions~\cite{sachdev2007quantum}. There has been a significant effort to simulate the ground state of many-body systems on classical computers. Semi-classical approaches, such as density functional theory~\cite{jones2015density}, are extensively used for characterizing the electronic structures but fail to explain all quantum effects. Quantum Monte Carlo~\cite{foulkes2001quantum} has been developed to go beyond such mean field approaches but they also suffer from sign problem. Density matrix renormalization group based methods~\cite{schollwock2005density}, are perhaps the most successful approach for characterizing the ground state features of many-body systems, although they are mostly limited to 1-dimensional systems. Machine learning~\cite{carrasquilla2017machine,carleo2017solving,gray2018machine,ch2017machine,wang2016discovering,nomura2017restricted} has also been exploited to capture the ground state features, but it is hard to scale them up for large systems. The limitations of these approaches truly reveals the limited power of classical simulations for unraveling quantum features. 

The emergence of quantum simulators has opened new vistas for simulating many-body systems. 
To date, there are three different methods for simulating the ground state of many-body systems on quantum simulators: (i) imaginary time evolution~\cite{motta2020determining, McArdle2019}; (ii) adiabatic evolution~\cite{10.1007/bf01343193,farooq2015adiabatic}; and (iii) Variational Quantum Eigensolver (VQE)~\cite{peruzzo2014variational}. The first two approaches are entirely performed on quantum simulators, and thus their success heavily relies on the quality of the quantum hardware. In particular, for large systems, these approaches either demand deep quantum circuits or long coherence times. However, the current Noisy Intermediate Scale Quantum (NISQ) simulators are prone to various types of error in their initialization, readout, and gate operations, while their finite coherence time limits both the system size and the depth of the circuit. To overcome these obstacles, the VQE method, a hybrid combination of a quantum circuit and a classical optimizer, has been theoretically investigated~\cite{10.22331/q-2019-10-07-191, 10.1088/1367-2630/18/2/023023, whitfield2011simulation, mcclean2017hybrid, barkoutsos2018quantum, romero2018strategies, Moll_2018, PhysRevA.92.042303} and experimentally implemented on various NISQ simulators, including photonic chips~\cite{peruzzo2014variational}, ion traps~\cite{shen2017quantum, hempel2018quantum, kokail2019self}, nuclear magnetic resonance systems~\cite{Li2011}, and superconducting quantum devices~\cite{o2016scalable, colless2018computation, kandala2017hardware}. 
In fact, the use of the classical optimizer allows one to simplify the quantum hardware into a shallow circuit. In other words, the resources needed for simulating the ground state is divided between quantum hardware and classical optimizer, enabling the imperfect NISQ simulators to emulate the ground state of many-body systems. 
Recently, VQE has been generalized to simulate excited states~\cite{Higgott2019variationalquantum, PhysRevA.99.062304}, non-equilibrium dynamics~\cite{li2017efficient}, Gibbs ensemble~\cite{wang2020variational}, and approximate solutions for combinatorial optimization problems~\cite{farhi2014quantum} in many-body systems. Based on these, it is highly desirable to have a quantitative analysis for the required resources, either quantum or classical, in different approaches for simulating the ground state of many-body systems (see e.g., Ref.~\cite{BravoPrieto2020scalingof} for resources in variational methods). 

In this paper, we use a digital quantum simulator to prepare the ground state of many-body Hamiltonians. We first compare the adiabatic evolution and the VQE with respect to their demand of quantum resources, namely the depth of the circuit or equivalently the number of two-qubit entangling gates. Our results quantitatively show that the VQE is far more efficient in terms of quantum resources than the adiabatic scheme. We then focus on the classical optimization part of the VQE to speed up its convergence and thus minimize the total time needed for the performance of the VQE. We specifically develop two different strategies for improving the initial guess for the circuit parameters to start the optimization at a closer point to the global minimum. Our analysis shows that these strategies can significantly speed up the convergence of the optimization, in particular, for larger systems. The results are very general and can be applied for a wide range of Hamiltonians even without symmetries.

\section{Model}\label{section:model}

Our goal is to prepare the ground state of a many-body system on a Digital Quantum Simulator (DQS). For simplicity, and without loss of generality, we consider a $1$-dimensional chain of $N$ spin-$1/2$ particles interacting via Heisenberg Hamiltonian
\begin{equation}
    H = J \sum_{i = 1}^{N-1} \boldsymbol{\sigma}^{i}\cdot\boldsymbol{\sigma}^{i + 1},
    \label{eq:Hamiltonian_Heis}
\end{equation}
where $J>0$ is the exchange coupling and $\boldsymbol{\sigma}^{i} = (\sigma_{x}^{i}, \sigma_{y}^{i}, \sigma_{z}^{i})$ is the vector of Pauli operators at site $i$. The Heisenberg model is one of the key models in both condensed matter and quantum information technologies, and thus, has been extensively explored in both ground state \cite{Mikeska2004} and non-equilibrium dynamics~\cite{endoh1974dynamics,bose2003quantum,de2006entanglement,bayat2010information}. 
The Heisenberg Hamiltonian has SU(2) symmetry and commutes with the total spin in the $\alpha=x,y,z$ direction (i.e. $[H, S_{\alpha}] = 0$ where $S_{\alpha} = \sum_{i} \sigma_{\alpha}^{i}$). In addition, the Hamiltonian also commutes with the total spin operator $S_{tot}^2=S_x^2+S_y^2+S_z^2$. This implies that every eigenstate of the system has a specific total spin $S_{tot}$ with a fixed $S_z$. In particular, for even $N$, assumed in this paper, the ground state $|GS\rangle$ is a global singlet with both $S_{tot} = 0$ and $S_z=0$. Furthermore, the Heisenberg Hamiltonian supports mirror-inversion symmetry as $[H,\mathcal{M}]=0$, where $\cal{M}$ is the mirror-inverting operator with $\mathcal{M} |i_1,i_2,\cdots,i_N\rangle = |i_N,\cdots,i_2,i_1\rangle$. This implies that the eigenstates of the system also follow the same symmetry. It is worth emphasizing that our methodology is easily generalized to other Hamiltonians, such as XXZ, which satisfy the conservation of $S_z$ and mirror-inversion symmetries. 
The ground state of the Heisenberg Hamiltonian has a matrix product state representation so that the results from quantum simulators can be certified by classical simulation. This allows us to enhance our confidence on the performance of quantum simulators, and paves the way for solving more complex problems which are not classically simulable, such as the ground state of $2$-dimensional systems.

\section{Digital Quantum Simulator}\label{section:DQS}

\begin{figure}[t]
    \centering
    \includegraphics[width=\columnwidth]{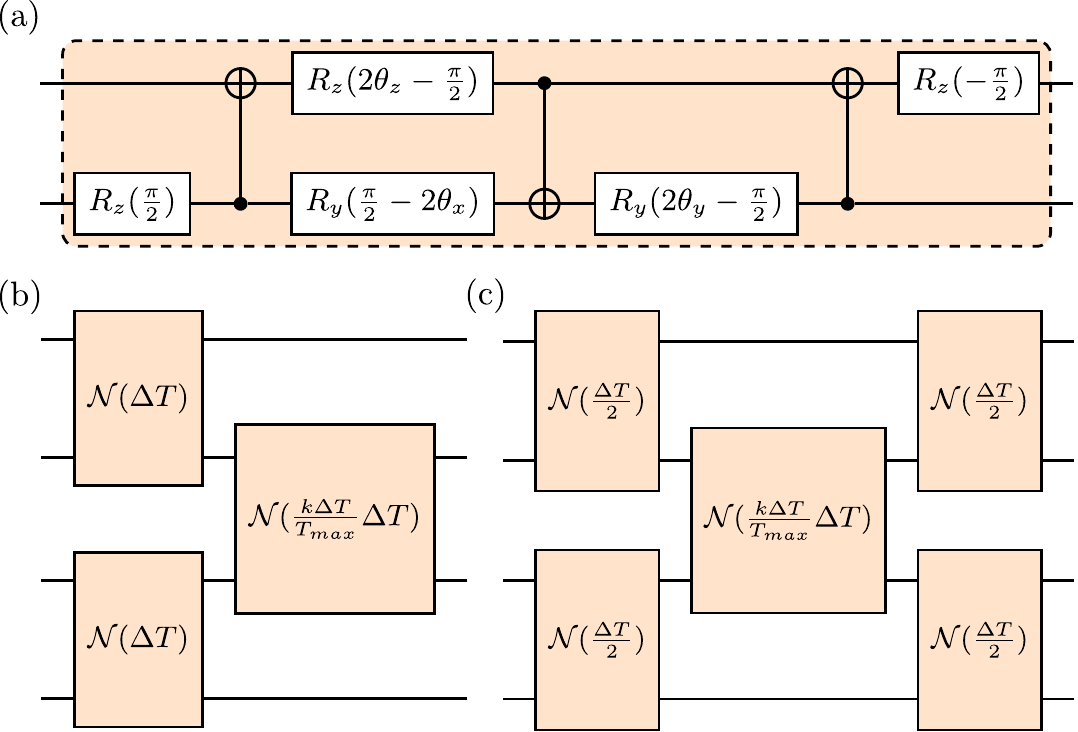}

    \caption{(a) Circuit for realizing $\mathcal{N}(\theta_x, \theta_y, \theta_z)$ as the entangling gate between two qubits. (b) A single layer circuit for realizing the first order Suzuki-Trotter approximation in a system of length $N=4$. (c) A single layer circuit for realizing the second order Suzuki-Trotter approximation in a system of length $N=4$.}
    \label{fig:circuits}
\end{figure}

A digital quantum simulator is a quantum machine which manipulates the quantum state of its qubits through performing certain one- and two-qubit gate operations. In order to be a universal quantum computer, a digital quantum simulator needs to be capable of applying arbitrary single qubit unitary operations (which can be decomposed into rotations around $x$, $y$ and $z$ axes) on all qubits together with at least one two-qubit entangling gate between any pair of particles~\cite{nielsen2002quantum}. In this paper, we consider all the single qubit unitary operations to be a combination of the following gates
\begin{equation}
P(\theta)=\begin{pmatrix}
1 & 0 \\
0 & e^{i\theta}
\end{pmatrix}, \text{ and  } R_\alpha(\theta)=e^{i\frac{\theta}{2} \sigma_\alpha}, \text{ with } \alpha=x,y,z\textbf{,}
\end{equation} 
where $P(\theta)$ is the phase gate and $R_\alpha(\theta)$ is the rotation around the $\alpha$-axis. For a two-qubit entangling gate we use controlled-$x$ (also known as CNOT) gate which is defined as 
\begin{equation}
	U_{CX}=\begin{pmatrix}
		1 & 0 & 0& 0 \\
		0 & 1 & 0& 0\\
		0 & 0 & 0& 1\\
		0 & 0 & 1& 0\\
	\end{pmatrix}.
\end{equation} 
By properly combining the above one- and two-qubit gates one can make any unitary operation. 
Nonetheless, there is not a straightforward recipe to design an efficient circuit to generate a specific target state. In particular, designing protocols for preparing an eigenstate of a Hamiltonian, on digital quantum simulators, has attracted significant attention in recent years~\cite{las2014digital,salathe2015digital,barends2016digitized,o2016scalable, colless2018computation, kandala2017hardware,hempel2018quantum}. We pursue two approaches for generating the ground state of many-body systems, namely adiabatic approach and VQE. 
In both methods, which will de discussed in the following sections, the entangling part of the circuit is a two-qubit unitary operator in the form of
\begin{equation}
	\mathcal{N}(\theta_x, \theta_y, \theta_z) = e^{i\left(\theta_x \sigma_{x} \otimes \sigma_{x} + \theta_y \sigma_{y} \otimes \sigma_{y} + \theta_z \sigma_{z} \otimes \sigma_{z}\right)}
	\label{eq:N(theta)}
\end{equation}
in which $\theta_{x,y,z}$ are three angles. One can actually implement this unitary operator using a simple circuit shown in Fig.~\ref{fig:circuits}(a), in which three CNOT gates are combined by five local rotations around $y$ and $z$ axes~\cite{vatan2004optimal}. Thanks to the symmetries of the Heisenberg Hamiltonian, throughout this paper, we consider $\theta_x{=}\theta_y{=}\theta_z$ and thus we omit the dependence on three independent parameters and simply use $\mathcal{N}(\theta)$. Note that by imposing the condition $\theta_x{=}\theta_y{=}\theta_z{=}\theta$, the operator $\mathcal{N}(\theta)$ not only conserves $S_{z}$, but also can create a maximally entangled state when acts on $| 01 \rangle$ for the choice of $\theta {=} \pi / 8$. Therefore, $\mathcal{N}(\theta)$ is truly a two-qubit entangling gate.

\section{Adiabatic Evolution}\label{section:Adiabatic}

Adiabatic theorem~\cite{10.1007/bf01343193} is a well-known procedure for the preparation of the ground state of a complex Hamiltonian starting from a simpler one. The system is initially prepared in the ground state of a simple Hamiltonian $H_{ad}(0)$ which is then slowly varied in time to a more complex desired Hamiltonian $H_{ad}(T_{max})$ after time $t=T_{max}$. If the variation of the Hamiltonian is slow enough, the quantum state of the system always remains in the ground state of the instant Hamiltonian $H_{ad}(t)$~\cite{nenciu1980adiabatic}. In this paper, we are focused in the preparation of the ground state of the Hamiltonian $H$ in Eq.~(\ref{eq:Hamiltonian_Heis}) using a digital quantum simulator. In Ref.~\cite{farooq2015adiabatic}, in an analog approach, a dimerized Heisenberg Hamiltonian is adiabatically evolved to the uniform Hamiltonian $H$ according to 
\begin{equation}
H_{ad}(t) = H_{odd} + \frac{t}{T_{max}} H_{even}
\end{equation} 
where $0 \le t \le T_{max}$ and   
\begin{equation}
H_{odd} = J\sum_{\text{odd } i} \boldsymbol{\sigma}^{i}\cdot\boldsymbol{\sigma}^{i + 1}, \text{ and     }
H_{even} = J\sum_{\text{even } i} \boldsymbol{\sigma}^{i}\cdot\boldsymbol{\sigma}^{i + 1}.
\end{equation} 
At $t=0$ the Hamiltonian is fully dimerized with only odd couplings switched on and the ground state of the system is simply $|\Psi(0)\rangle=|\psi^-\rangle \otimes \cdots \otimes |\psi^-\rangle$, where $|\psi^-\rangle=(|01\rangle-|10\rangle)/\sqrt{2}$. By choosing $T_{max}\sim N^2$ the condition of adiabatic theorem~\cite{nenciu1980adiabatic} is satisfied and at $t=T_{max}$, where $H_{ad}(T_{max})=H$, the quantum state of the system becomes the ground state of the desired Hamiltonian $H$~\cite{farooq2015adiabatic}. 

To obtain the time evolution of the system, governed by $H_{ad}(t)$, on a digital quantum simulator one has to adopt two essential steps: (i) discretize the time evolution into $M_{ad}$ steps over which the Hamiltonian remains fixed for a time interval of $\Delta t = T_{max}/M_{ad}$; and (ii) exploit the Suzuki-Trotter expansion for evolving the system in each time step. In each discretized time step $k$ (with $k$ going from $1$ to $M_{ad}$) the Hamiltonian is considered to be fixed, namely $H_{ad}(k\Delta t)$. The time evolution operator at time step $k$ is thus written as
\begin{equation} \label{U_adiab_Delta_t}
U_k(\Delta t)= e^{-i H_{ad}(k\Delta t) \Delta t} = e^{-i (H_{odd}+\frac{k\Delta t}{T_{max}} H_{even}) \Delta t }.
\end{equation}   
Notice that, in order to emulate the true time evolution of the $H_{ad}(t)$ the discrete time steps have to be small or equivalently the number of steps $M_{ad}$ has to be large. This will be discussed in more details in the following. As mentioned above, the second essential step for realizing the time evolution of a many-body system on a digital quantum simulator is to use Suzuki-Trotter expansion~\cite{hatano2005finding}. This will allow us to simulate the dynamics only through one- and two-qubit gate operations. We consider both the first and the second order realization of the Suzuki-Trotter expansion for the unitary operator $U_k(\Delta t)$, given in Eq.~(\ref{U_adiab_Delta_t}), at the time step $k$ , namely   
\begin{eqnarray}
U_k^{\text{ST1}}(\Delta t) &=& e^{-iH_{even} \frac{k\Delta t^2}{T_{max}} } e^{-iH_{odd} \Delta t},  \cr
U_k^{\text{ST2}}(\Delta t) &=& e^{-iH_{odd} \frac{\Delta t}{2}} e^{-iH_{even} \frac{k\Delta t^2}{T_{max}} } e^{-iH_{odd} \frac{\Delta t}{2}}.
\end{eqnarray}
While the first order Suzuki-Trotter operator $U_k^{\text{ST1}}(\Delta t)$ approximates the discrete time evolution $U_k(\Delta t)$ by a quadratic error as  $U_k(\Delta t)\approx U_k^{\text{ST1}}(\Delta t)+O(\Delta t^2)$, the second order Suzuki-Trotter operator $U_k^{\text{ST2}}(\Delta t)$ improves the approximation as $U_k(\Delta t)\approx U_k^{\text{ST2}}(\Delta t)+O(\Delta t^3)$. In Figs.~\ref{fig:circuits}(b)-(c) we present the quantum circuit for each time step of the first and the second order Suzuki-Trotter evolution, respectively. 

\begin{figure}[t]
    \centering
    \includegraphics[width=\columnwidth]{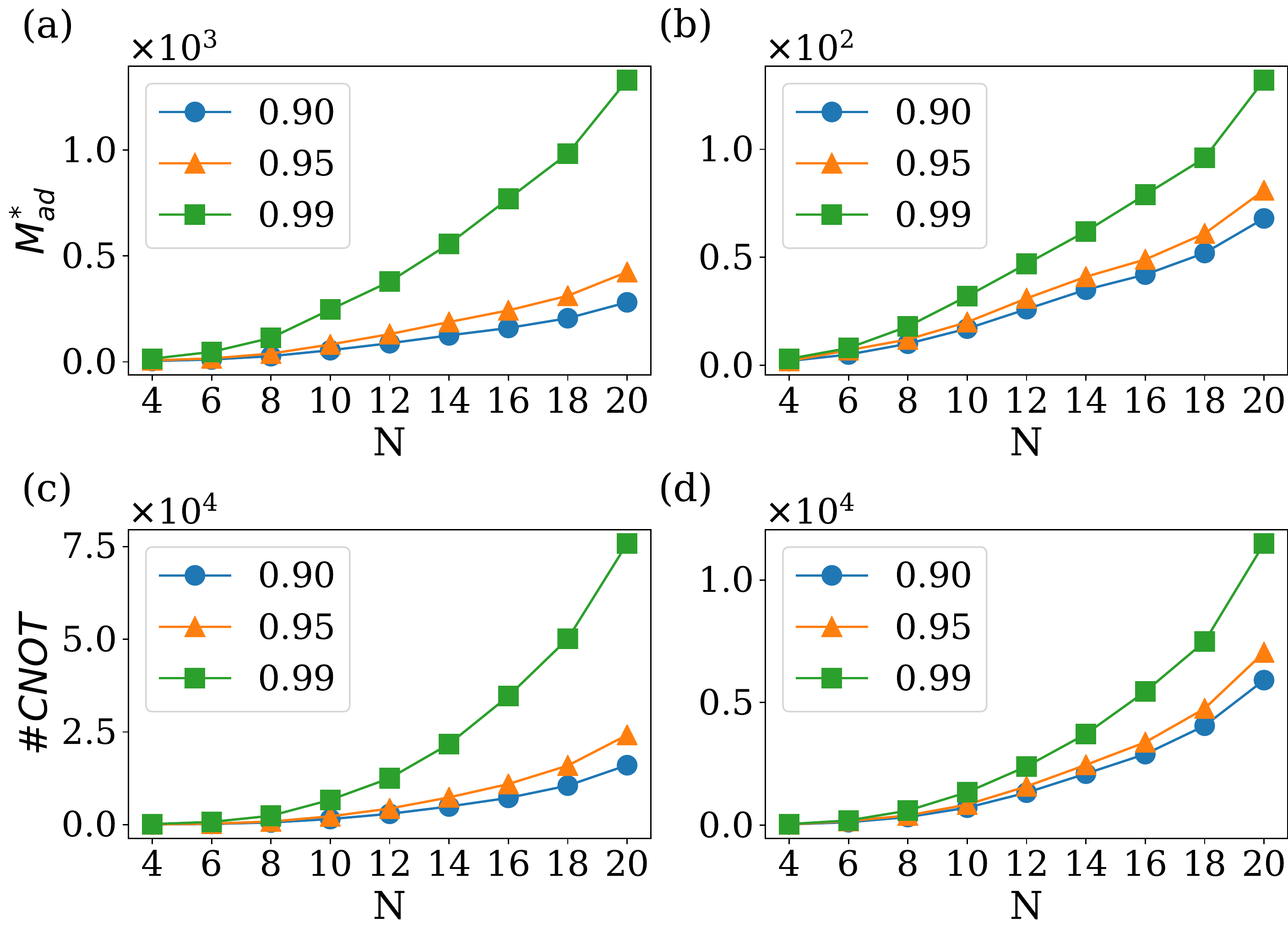}
	\caption{The minimum circuit depth $M_{ad}^*$ as a function of $N$ for various threshold fidelities using: (a) first order Suzuki-Trotter approximation; and (b)  Second order Suzuki-Trotter procedure. The number of CNOT gates required for the adiabatic evolution as a function of length $N$ for required for various threshold fidelities using: (c) first order Suzuki-Trotter approximation; and (d)  Second order Suzuki-Trotter procedure.}
	\label{fig:ad_result}
\end{figure}

In order to speed up the adiabatic protocol, we first determine the minimum time $T_{max}$ which is required to prepare the system in its ground state for a given threshold fidelity $F=|\langle \Psi(T_{max})|\text{GS}\rangle|^2$ between the quantum state of the system $|\Psi(T_{max})\rangle$ and the target state $|\text{GS}\rangle$. By fixing the threshold fidelity $F$ and system size $N$, one can solve the Schr\"{o}dinger equation with time dependent Hamiltonian $H_{ad}(t)$ and find the minimum $T_{max}$ which results in a fidelity above the threshold $F$. As shown in Ref.~\cite{farooq2015adiabatic}, the minimum time needed scales as $T_{max}\sim N^2$ where the proportionality coefficient gets larger as the threshold fidelity increases.  By specifying the minimum value of $T_{max}$ one can then apply the above adiabatic protocol on a digital quantum simulator to find the minimum required Suzuki-Trotter steps $M_{ad}^*$ to achieve the threshold fidelity $F$. In Figs.~\ref{fig:ad_result}(a)-(b), we plot $M_{ad}^*$ as a function of $N$ for various threshold fidelities when the first and the second order Suzuki-Trotter circuits are used respectively. Notably, for a given threshold fidelity, the number of layers in the the second order Suzuki-Trotter approximation is almost one order of magnitude smaller than the the first order one. However, the number of gates in each layer of the second order Suzuki-Trotter circuits are more than the number of gates in their corresponding first order. Hence, it is more meaningful to compare the number of gates needed for achieving the threshold fidelity rather than the number of layers. In practice, the two-qubit gates are far more challenging than the one-qubit gates. Therefore, in Figs.~\ref{fig:ad_result}(c)-(d) we plot the number of CNOTS versus the system size $N$ for various threshold fidelities in both first and second order Suzuki-Trotter circuits, respectively. Remarkably, the second order Suzuki-Trotter circuit shows a clear superiority over the first order by demanding significantly less number of CNOT gates for delivering the same fidelity. This clearly shows that the second order Suzuki-Trotter expansion is significantly more resource efficient than the first order for realizing the ground state of a many-body system on a digital quantum simulator.

\section{Variational Method}

As shown in the previous section, the adiabatic approach for creating the ground state of many-body systems demands deep circuits, i.e. large $M_{ad}^*$, with a considerable number of gates. The near-term quantum simulators cannot provide such deep circuits due to their imperfect noisy quantum gates, in particular the two-qubit gates. Therefore, it is worthy of exploring the possibilities for the use of shallow circuits to simulate the ground state of many-body systems. Variational methods, such as variational quantum eigensolver (VQE) algorithm, provide an alternative to adiabatic evolution which can be realized in shallow circuits. In these methods, the post processed measurement outcomes of a fairly shallow quantum circuit has to be fed into a classical optimizer iteratively to find the optimal circuit parameters for realizing the ground state of a many-body system. In the VQE algorithm, the quantum device prepares a quantum state $|\psi(\vec{\theta})\rangle$, which is often called the ansatz, with $\vec{\theta}=(\theta_1,\theta_2,\cdots,\theta_L)$ being some controllable parameters in the circuit. Then one can measure the average energy $\langle H \rangle= \langle\psi(\vec{\theta})|H|\psi(\vec{\theta})\rangle$ for this quantum state through appropriate measurements on the qubits of the simulator. For the Heisenberg model in Eq.~(\ref{eq:Hamiltonian_Heis}), this reduces to measure correlation functions between adjacent qubits and then add them up, namely $\langle H \rangle=\sum_{i=1}^{N-1} \langle\psi(\vec{\theta})| \boldsymbol{\sigma}^{i}\cdot\boldsymbol{\sigma}^{i + 1} |\psi(\vec{\theta})\rangle$. The measured average energy can be used as an input for a classical optimizer to find its minimum by adaptively adjusting the parameters $\vec{\theta}$ on the simulator. The minimization algorithm eventually finds the optimal parameters $\vec{\theta}_{min}$ for which $|\psi(\vec{\theta}_{min})\rangle$ is the ground state with minimum average energy. Therefore, the resources required for VQE is divided between the quantum simulator and the classical optimizer. In fact, in each Classical Optimization Iteration (COI) a new experiment on the quantum simulator is needed, making the VQE very time consuming. Thus, for the success of the VQE, two steps are essential: (i) a quantum circuit, as shallow as possible with minimal number of parameters; and (ii) an efficient classical optimizer which can find the minimum of average energy using least number of iterations.     

\begin{figure*}
    \centering

    \includegraphics[width=\textwidth]{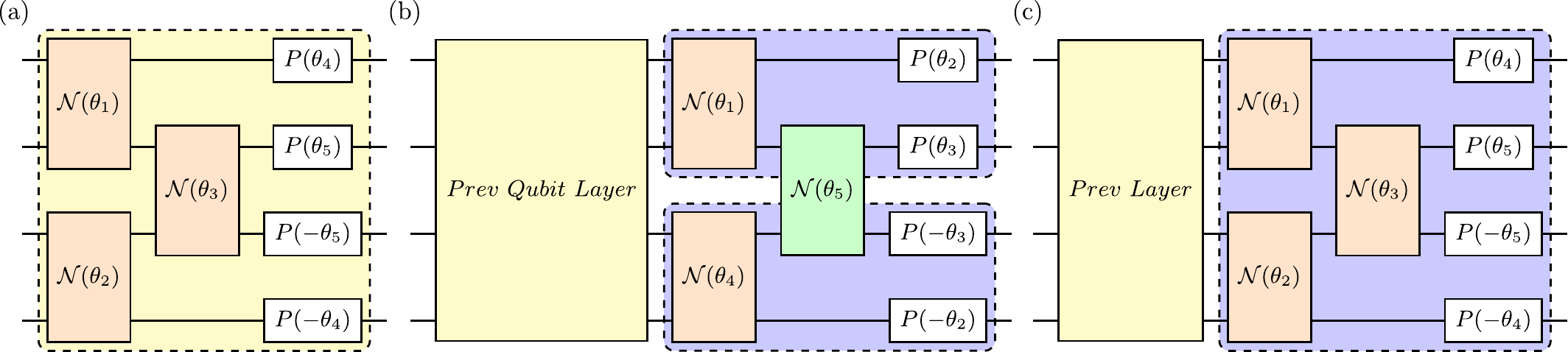}

    \caption{(a) The schematic picture of a single layer VQE circuit for a chain of length $N=4$. (b) The qubit recursive algorithm for creating a large circuit using the optimized parameters of two subsystem with smaller sizes, here half size. Apart from the optimized gates, one entangling gate, shown in green, initialized by a random number is added between the two subsystems for connecting them. If the larger system needs more layers than the smaller ones then the extra layers are initialized with the parameters copied from the previous layer. (c) The layer recursive algorithm for making the circuit deeper step by step after optimizing each layer. The optimized parameters of the previous layers are used as the initial guess for the new circuit when a new layer is added. The parameters in the new layer are copied from the previous one.}
    \label{fig:ansatz}
\end{figure*}

Designing a quantum circuit, with minimal resources, which is capable of realizing the ground state of the Hamiltonian, here Heisenberg interaction, is a key step in a VQE algorithm~\cite{herasymenko2019diagrammatic}. The output of the circuit will be $|\psi(\vec{\theta})\rangle$ and by varying the parameters $\vec{\theta}$ one can explore the Hilbert space for finding the quantum state with minimum average energy. Ideally, the quantum circuit should be able to explore the whole Hilbert space as $\vec{\theta}$ vary. However, using symmetries of the system one can simplify the circuit, i.e. decreasing the number of parameters in $\vec{\theta}$, and instead of exploring the whole Hilbert space only focus on the relevant part of the Hilbert space. 
For instance, in the case of the Heisenberg Hamiltonian, considering the preservation of $S_z$ and the mirror-inversion symmetries can significantly reduce the number of parameters in $\vec{\theta}$ which not only simplifies the quantum circuit but also makes the convergence of the classical optimization procedure faster. Considering these two symmetries, we suggest to use $\mathcal{N}(\theta)$, introduced in Eq.~(\ref{eq:N(theta)}), as the  entangling gates of the circuit because they do not create any extra excitation in the system. These gates can be combined in a geometry similar to the first order Suzuki-Trotter steps. In order to be able to locally manipulate the phases of each qubit we can also add phase gates to every qubit of the circuit as shown in Fig.~\ref{fig:ansatz}(a). The mirror-inversion symmetry implies that for even (odd) $N$ the angles of the phase gates acting on site $k$ and $N-k+1$ are equal with the opposite (the same) signs. Therefore, one can easily show that the total number of parameters $L$ in a VQE circuit is
\begin{equation}\label{eq:num_param_VQE}
L = \left(\frac{3N}{2} - 1\right) M_{VQE}
\end{equation} 
where, $M_{VQE}$ represents the number of layers. The proposed circuit is capable to explore a part of Hilbert space which includes the ground state. In principle, one can chose the parameters such that this circuit replicates the adiabatic circuit in Fig.~\ref{fig:circuits}(b)-(c), e.g. by putting the angles of the phase gates to zero. However, in the adiabatic circuit every layer is close to identity operator as it evolves the system only for small time $\Delta t$. This results in a deep circuit demanding a considerable number of gates as shown in Fig.~\ref{fig:ad_result}. The idea behind the VQE is to exploit an optimization algorithm to find the optimal set of parameters for which a shallow circuit can implement the ground state of a many-body system. 

\setlength{\tabcolsep}{0.23em} 
\begin{table}
    \centering
    \begin{tabular}{|c|c|c|c|c|c|c|}
        \hline
        & \multicolumn{2}{|c|}{$ST~1st~order$} & \multicolumn{2}{|c|}{$ST~2nd~order$} & \multicolumn{2}{|c|}{$VQE$} \\
        \hline
        N & $M^{*}_{ad}$ & $\#CNOT$ & $M^{*}_{ad}$ & $\#CNOT$ & $M^{*}_{VQE}$ & $\#CNOT$ \\
        \hline
        4 & 15 & 135 & 3 & 45 & 2 & 18 \\
        \hline
        8 & 113 & 2373 & 18 & 594 & 3 & 63 \\
        \hline
        10 & 247 & 6669 & 32 & 1344 & 3 & 81 \\
        \hline
        16 & 770 & 34650 & 79 & 5451 & 5 & 225 \\
        \hline
        20 & 1330 & 75810 & 132 & 11484 & 6 & 342 \\
        \hline
    \end{tabular}
    \caption{A comparison for the circuit depth $M^*$ between the adiabatic approach (with both first and second order Suzuki-Trotter approximation) and the VQE algorithm for various system sizes when the threshold fidelity is chosen to be $F=0.99$.}
    \label{tbl:layers}
\end{table}

In order to finalize the VQE circuit, apart from designing each layer, one has to also determine the number of layers $M_{VQE}$. To find out the number of layers needed to successfully produce the ground state, one can use the fidelity between the quantum state of the circuit and the real ground state at the end of optimization procedure. The layers are increased one by one so that the attainable fidelity exceeds a threshold value. Note that fidelity is not an experimentally friendly quantity as one needs to know the quantum state. Therefore, in practice, instead of fidelity one can focus on the convergence of optimized average energy as the number of layers increases. 
For any system size $N$ and a chosen value for the layer number $M_{VQE}$ we use $50L$ (for $L$ being the number of parameters in $\vec{\theta}$) COI to see whether the obtained fidelity (or average energy) reaches the threshold value. If the optimization fails to reach the threshold fidelity then we increase the layer number $M_{VQE}$ by one unit until the algorithm is successful.  

In every circuit, we initially start with a random guess for all the parameters in $\vec{\theta}$ and use Adam algorithm (with amsgrad applied)~\cite{kingma2014adam, reddi2019convergence}, as an extension to stochastic gradient descent, to minimize our cost function, namely the average energy of the system. In Adam algorithm there are four different hyper parameters which have to be set as the input of the optimizer: (i) the learning rate $\alpha$, which controls how fast the parameters are going to be updated in each iteration. (ii) $\beta_{1}$, the exponential decay rate for the first moment estimation; (iii) $\beta_{2}$, the exponential decay rate for the second-moment estimation; and (iv) a very small number $\epsilon$ to prevent any division by zero. We set the hyper parameters to be $\alpha=0.01$, $\beta_{1}=0.9$, $\beta_{2}=0.999$ and $\epsilon=10^{-8}$.

In order to be independent of the initial random guess for the parameters $\vec{\theta}$ we repeat the procedure for several ($\sim 100$) different random samples. Only if the fidelity of all of them exceeds the threshold after $50L$ iterations we call it a success.
For our numerical simulations, we use a cluster equipped with 192 cpu cores and 512~GB memory. For the largest system size considered here $N=20$, the simulations take nearly a week to finish 100 different random samples using such hardware.
Similar to the adiabatic approach, the minimum number of layers which can successfully generate the ground state within the threshold fidelity is called $M_{VQE}^*$. In Table.~\ref{tbl:layers}, we compare $M_{VQE}^*$ with $M_{ad}^*$ in both first and second order Suzuki-Trotter approximation and their corresponding CNOT gate numbers for different system sizes when the threshold fidelity is chosen to be $F=0.99$. As it is clear from the Table, the number of required layers and the CNOT gate numbers in the VQE approach is significantly smaller than the adiabatic procedure. Even for a system of size $N=20$, a fairly shallow circuit of $M_{VQE}^*=6$ layers suffices to produce the ground state of the Heisenberg chain with the fidelity exceeding $F=0.99$. This remarkable observation shows the superiority of the VQE over the adiabatic approach with respect to the complexity of the circuit. The simplification in quantum resources, i.e. the circuit, is achieved thanks to the employment of a classical optimizer. Thus, in order to truly investigate the complexity of the VQE approach, one has to consider the required resources for the classical optimization procedure as well.

\begin{figure}[t]
    \centering
    \includegraphics[width=\columnwidth]{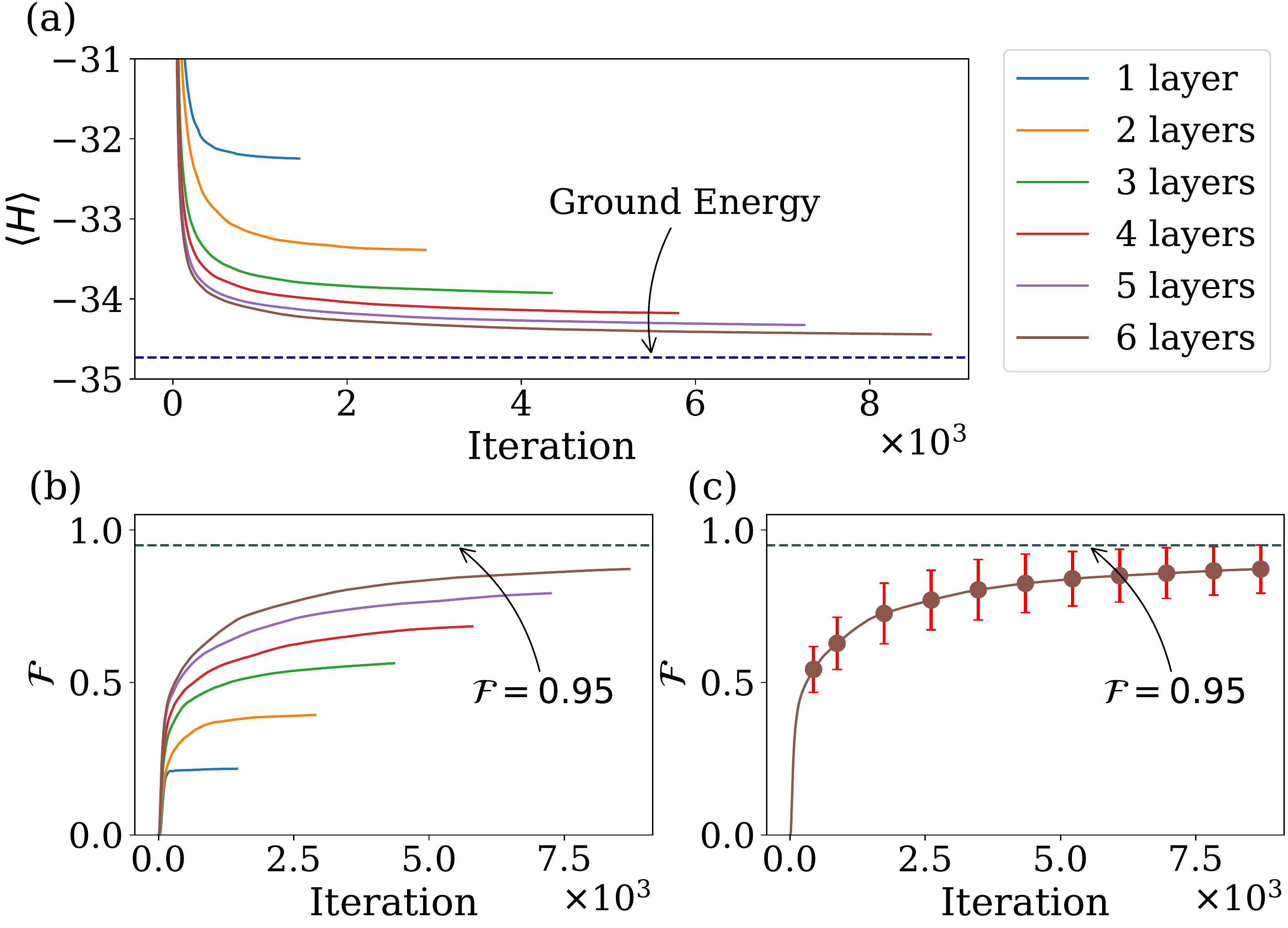}

    \caption{(a) The average energy as a function of COI in random strategy (i.e. strategy 1) for various circuit layers in a chain of length $N=20$. As the circuit layer increases the optimized average energy approaches the real ground state energy. (b) The obtainable fidelity averaged over 100 different random samples as a function of COI for circuits of different depth. (c) The average obtainable fidelity, and its corresponding error bars, as a function of COI when the whole VQE algorithm is repeated for $100$ times. }
    \label{fig:random}
\end{figure}

\begin{figure}[t]
    \centering
    \includegraphics[width=\columnwidth]{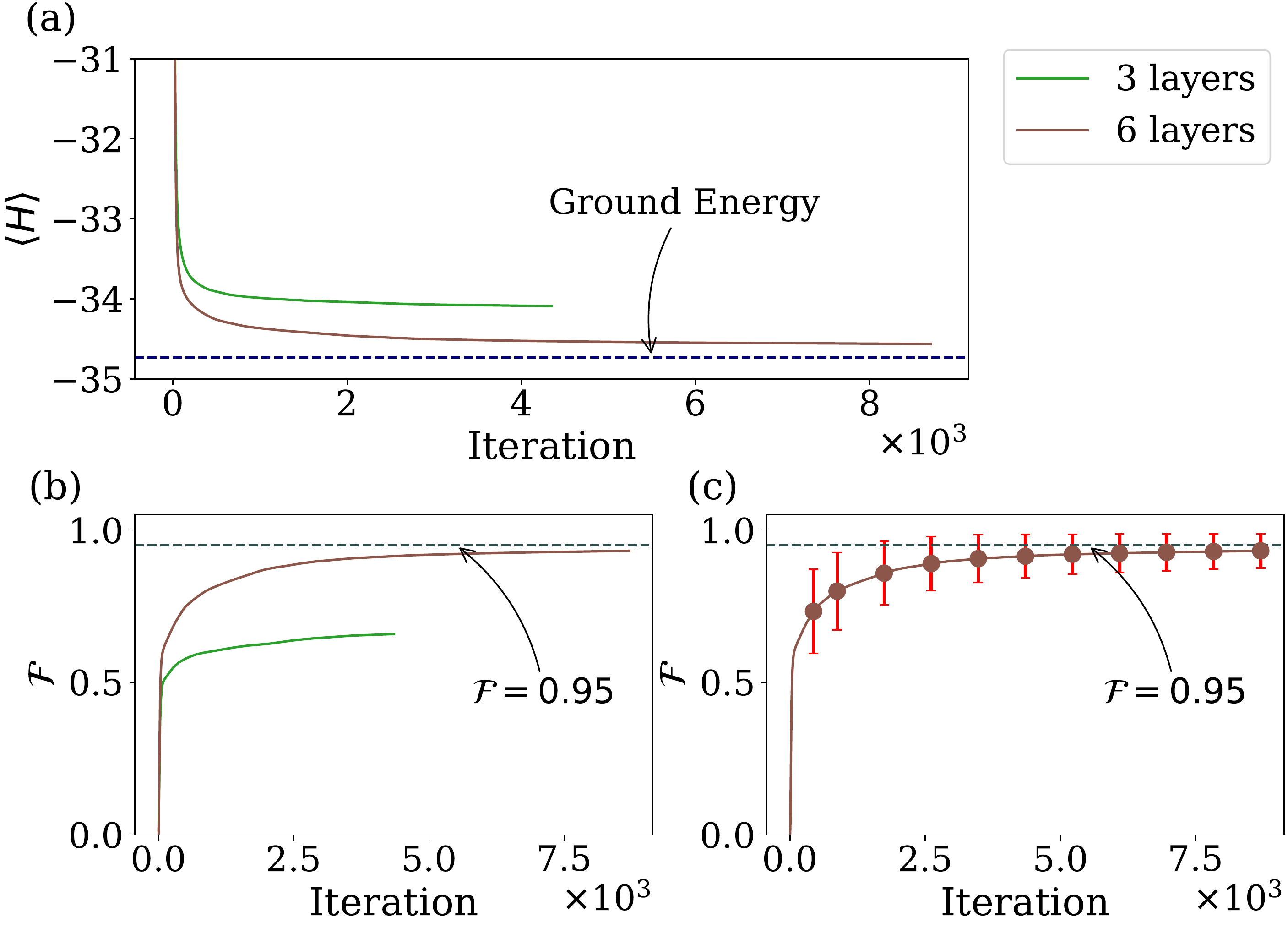}
    \caption{(a) The average energy as a function of COI in qubit recursive strategy (i.e. strategy 2) for various circuit layers in a chain of length $N=20$. As the circuit layer increases the optimized average energy approaches the real ground state energy. (b) The obtainable fidelity averaged over 100 different random samples as a function of COI for circuits of different depth. (c) The average obtainable fidelity, and its corresponding error bars, as a function of COI when the whole VQE algorithm is repeated for $100$ times.}
    \label{fig:qubit}
\end{figure}

\begin{figure}[t]
    \centering
    \includegraphics[width=\columnwidth]{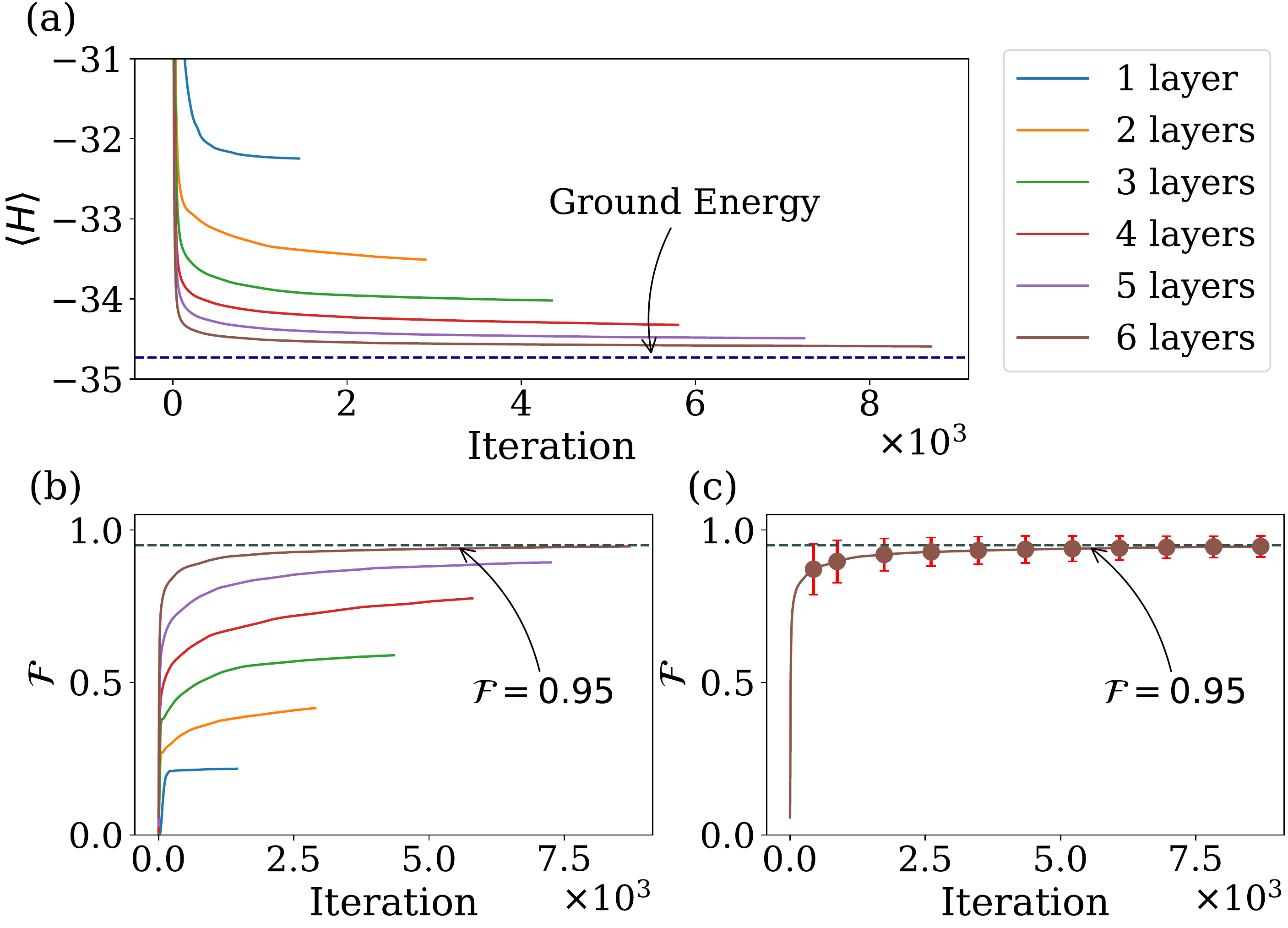}

    \caption{(a) The average energy as a function of COI in layer recursive strategy (i.e. strategy 3) for various circuit layers in a chain of length $N=20$. As the circuit layer increases the optimized average energy approaches the real ground state energy. (b) The obtainable fidelity averaged over 100 different random samples as a function of COI for circuits of different depth. (c) The average obtainable fidelity, and its corresponding error bars, as a function of COI when the whole VQE algorithm is repeated for $100$ times.}
    \label{fig:layer}
\end{figure}

\section{Strategies for accelerating the classical optimization}

A conventional classical optimization procedure starts with a random guess for $\vec{\theta}$ and iteratively minimizes the cost function, here the average energy. However, the random initial guess can be very far from the real minimum of the cost function. This may lead to two unwanted outcomes: (i) many optimization iterations are needed to get to the global minimum; and (ii) it is likely that the optimization gets trapped in a local minimum and thus fails to reach the right answer. Therefore, it is of high interest to see whether one can start the optimization with a smart guess which is relatively close to the real minimum of the cost function. In the following, we first explain the random initial guess in details, as the first strategy, and then provide two  different strategies with which we can speed up the convergence and improve the precision. \\

\textbf{Strategy 1: random initialization.} We first provide a detailed introduction for the basic random initialization strategy. A single circuit layer which is used in this strategy is shown in Fig.~\ref{fig:ansatz}(a). One has to cascade several layers in order to achieve a high quality output. In a circuit of $N$ qubits and $M_{VQE}$ layers there are $L$ number of parameters, which is given in Eq.~(\ref{eq:num_param_VQE}). In this strategy, the initial values of the parameters are sampled from a Gaussian distribution of mean $0$ and variance $1$. In Fig.~\ref{fig:random}(a) we plot the average energy as a function of optimization iterations for circuit of different layers. As the figure shows by increasing the iterations the average energy decreases steadily, however, if the number of layers are small the final converged value is far from the ground state energy. This shows that if the circuit is too shallow then it has no capacity to reach the ground state merely by classical optimization. As the number of layers increases the converged value is closer to the real ground state energy. In the Fig.~\ref{fig:random}(b) we give the corresponding obtainable fidelity averaged over 100 different random samples as a function of COI for circuits of
different circuit layers $M_{VQE}$. As the average energy approaches the real ground state the fidelity also goes to $1$. In order to see the stability of the protocol, we repeat the optimization for 100 different random samples. In Fig.~\ref{fig:random}(c) we plot the average of obtained fidelities together with the error bars, given by the standard deviation, as a function of iterations for a system of length $N=20$ and a circuit with $M_{VQE}=6$ layers. Even after $\sim 7500$ iterations the average fidelity does not reach $0.95$ showing that the random initial choice is not an efficient guess for the optimizer.

\textbf{Strategy 2: qubit recursive.} The first strategy to improve the classical optimization procedure is to use the optimized value of a smaller system as the initial guess for the larger one. Here, we focus on the case that we use the optimized parameters in a circuit of size $N/2$ as the initial guess for a system of size $N$. Let's first consider the case that the number of layers $M_{VQE}$ is identical for both the systems of size $N/2$ and $N$. In this case, we simply replicate the optimized circuit of size $N/2$ twice to make a larger system of size $N$. To have a complete circuit one has to add one entangling gate $\mathcal{N}(\theta)$ to each layer at the connection between the two copies acting on odd couplings, as shown in Fig.~\ref{fig:ansatz}(b). Therefore, in total, one has to add $M_{VQE}$ different entangling gates between the copies which are initialized with random guesses. Apart from these newly added gates, the rest of the one- and two-qubit gates are all initialized with the optimized value from the system of size $N/2$. The motivation behind this choice is that, the reduced density matrix of the subsystems from the ground state of the large system has large overlap with the ground state of the local system. Thus, using this strategy, one expects to converge to the real global minimum faster. 
In the case that the number of layers in the small chain is less than the layers required for the large system we simply repeat the same parameters of the last layer for the added layers.  
The convergence of the average energy versus the COI is shown in Fig.~\ref{fig:qubit}(a) for different circuit layers. In Fig.~\ref{fig:qubit}(b) the corresponding obtainable fidelity averaged over 100 different random samples are depicted as a function of COI. Indeed, the results show faster convergence for each choice of layers.
In order to see the overall performance, we repeat the optimization for 100 different initial random guesses in a system of size $N=20$ with $M_{VQE}=6$ layers. The average obtained fidelities, with corresponding error bars, are plotted in Fig.~\ref{fig:qubit}(c) as a function of iterations. The performance shows clear improvement in comparison with the random strategy as the average fidelity exceeds the threshold fidelity of $F=0.95$ only after $\sim 3500$ iterations. 

\textbf{Strategy 3: layer recursive.} In this strategy, we start a circuit with only one layer. Obviously this circuit is too shallow to provide the ground state for large systems. Nonetheless, one can perform the optimization using random initial guess for minimizing the average energy. The minimization converges normally fairly quick, as the circuit is shallow and the number of parameters are not that large. However, the output state may have a small fidelity with the real ground state as the circuit is shallow. Hence, in the next step we add one more layer, which means that the circuit becomes deeper and more parameters exist to be optimized. The initial values for the parameters of the newly added layer are copied from the previous layer. At the end of the optimization, one can add a new layer and repeat this procedure. A schematic of the procedure is shown in Fig.~\ref{fig:ansatz}(c). In Fig.~\ref{fig:layer}(a) we plot the  average energy versus the number of iterations for different layers. As the figure shows while the energy converges by increasing the iterations the final value reaches the ground state energy only for circuits with adequate depth. In order to see the overall performance, we repeat the procedure $100$ times and compute the final fidelities. In Fig.~\ref{fig:layer}(b), we illustrate the obtainable fidelity averaged over 100 different random samples as a function of COI. As evident from the figure, this strategy shows faster convergence as contrasted with the previous ones. In Fig.~\ref{fig:layer}(c) we plot the average fidelities, with the corresponding error bar, as a function of COI in a circuit of size $N=20$ and depth $M=6$. As the figure shows the speed of convergence is sensibly faster than the random initial guess, discussed in the strategy 1, and the average fidelity exceeds the threshold fidelity $F=0.95$ only after $\sim 2500$ iterations. \\

\setlength{\tabcolsep}{0.19em}
\begin{table}[t]
    \centering
    \begin{tabular}{|c|c|c|c|}
        \hline
        & 1000 iter. & 3000 iter. & 6000 iter. \\
        \hline
        Random & 0.648 $\pm$ 0.090 & 0.786 $\pm$ 0.098 & 0.849 $\pm$ 0.087 \\
        \hline
        Qubit & 0.810 $\pm$ 0.123 & 0.898 $\pm$ 0.084 & 0.923 $\pm$ 0.064 \\
        \hline
        Layer & 0.916 $\pm$ 0.049 & 0.939 $\pm$ 0.024 & 0.947 $\pm$ 0.023 \\
        \hline
    \end{tabular}
    \caption{ Comparing the obtainable fidelities between the three different VQE strategies for given number of iterations in a system of size $N=20$.  }
    \label{tbl:fidelity_comparison}
\end{table}

In order to compare the three VQE strategies, in Table~\ref{tbl:fidelity_comparison}, we report the average fidelities obtained in each strategy for three different iterations in a system of size $N=16$ and a circuit of $M=6$ layers. The data clearly shows  significant improvement of both qubit and layer recursive strategies over the random one, in particular, when the number of iterations are smaller. Interestingly, the layer recursive strategy also shows modest superiority over the qubit recursive strategy. This means that optimizing the system layer by layer and use their optimized data as the initial guess for building a deeper circuit significantly enhances the convergence of the classical optimization in a VQE algorithm. This is the main result of our paper.

\begin{figure}[t]
    \centering
    \includegraphics[width=\columnwidth]{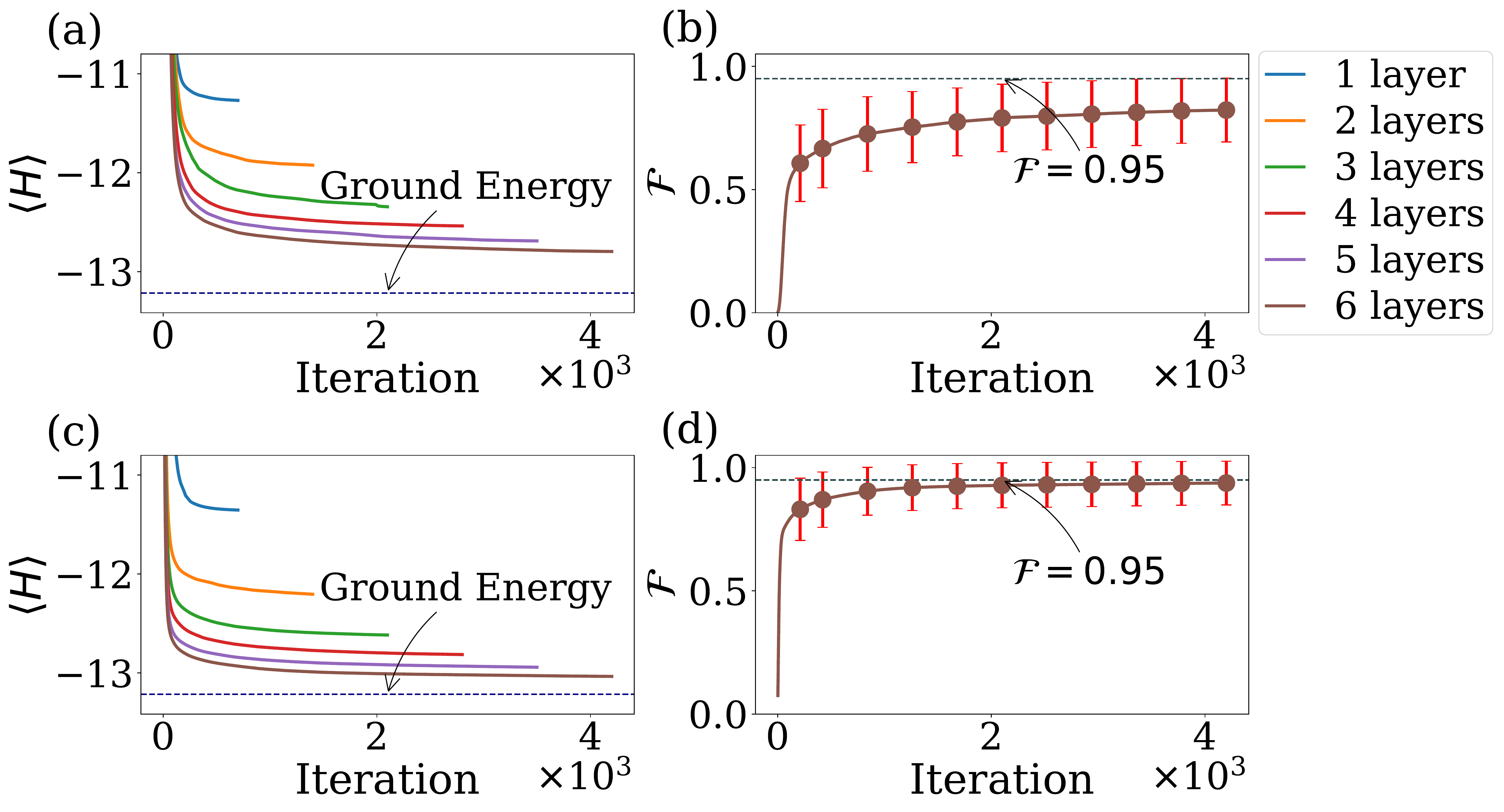}
    \caption{Simulating the ground state of a XYZ Hamiltonian using the mirror symmetry rule. (a) The average energy as a function of COI in random initialization strategy (i.e. strategy 1) for various circuit layers in a chain of length $N=10$. As the circuit layer increases the optimized average energy approaches the real ground state. (b) The average obtainable fidelity, and its corresponding error bars, as a function of COI when the whole VQE algorithm is repeated for $100$ times in random initialization strategy. (c) The average energy as a function of COI in layer recursive strategy (i.e. strategy 3) for a chain of $N=10$. As the circuit layer increases the optimized average energy approaches the real ground state faster than in strategy 1. (d) The average obtainable fidelity, and its corresponding error bars, as a function of COI when the whole VQE algorithm is repeated for $100$ times in layer recursive strategy.}
    \label{fig:XYZ_symmetry_applied}
\end{figure}

\begin{figure}[t]
    \centering
    \includegraphics[width=\columnwidth]{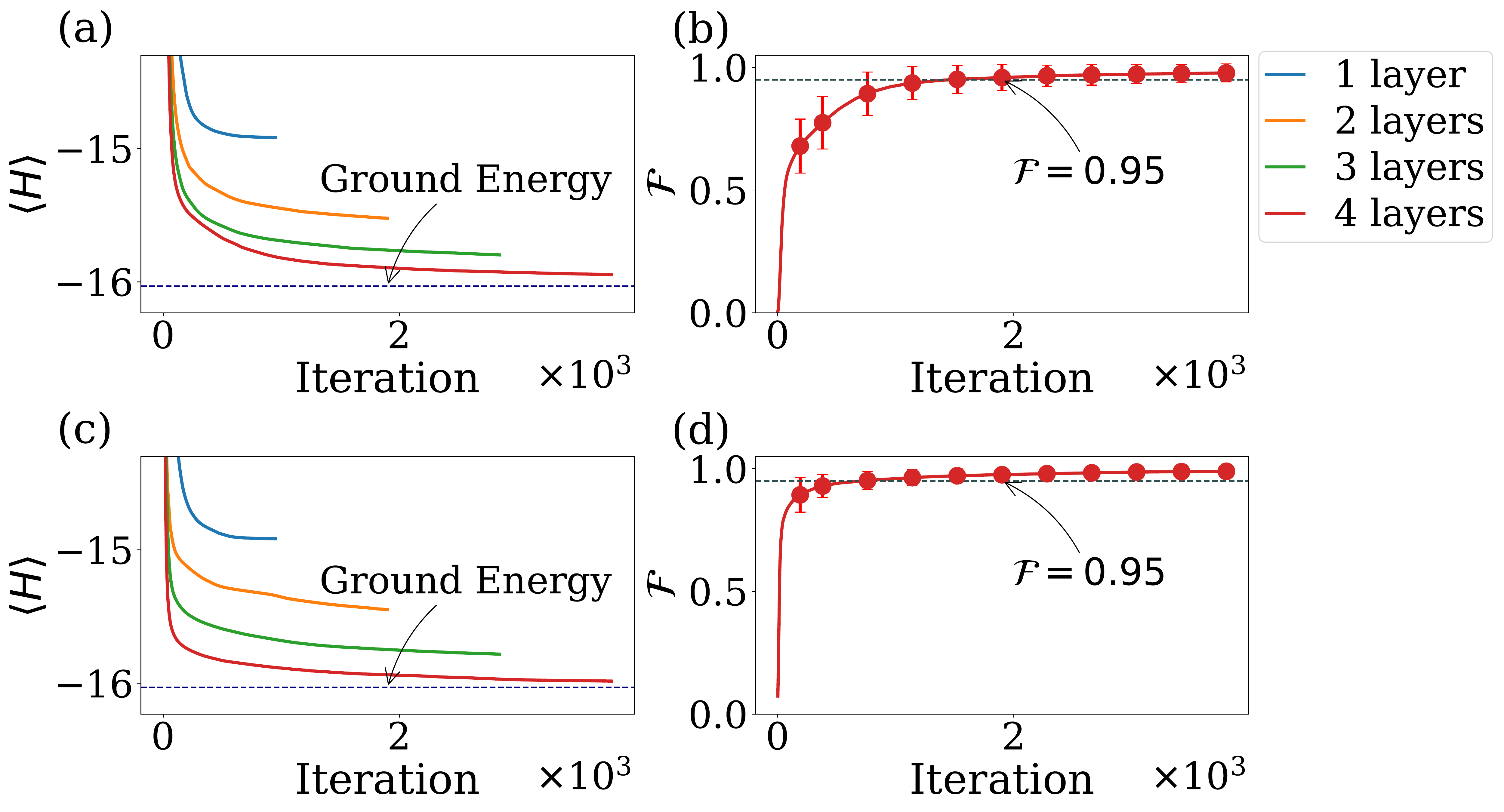}
    \caption{Simulating the ground state for a spin chain Kondo Hamiltonian. (a) The average energy as a function of COI in random initialization strategy (i.e. strategy 1) for various circuit layers in a chain of length $N=10$. As the circuit layer increases the optimized average energy approaches the real ground state energy. (b) The average obtainable fidelity, and its corresponding error bars, as a function of COI when the whole VQE algorithm is repeated for $100$ times in random initialization strategy. (c) The average energy as a function of COI in layer recursive strategy (i.e. strategy 3) for various circuit layers in a chain of length $N=10$. As the circuit layer increases the optimized average energy approaches the real ground state energy faster than in the strategy 1. (d) The average obtainable fidelity, and its corresponding error bars, as a function of COI when the whole VQE algorithm is repeated for $100$ times in layer recursive strategy.}
    \label{fig:kondo}
\end{figure}

\section{Generality of the protocol}

In the previous sections, we have focused on the Heisenberg Hamiltonian for which we have exploited two major symmetries of the system, namely the conservation of $S_{z}$ and the mirror symmetry. Here, we generalize our analysis and show that our protocol and the developed accelerated algorithms work well for more general Hamiltonians where these symmetries are not applicable. As the first example, we consider the general XYZ Hamiltonian 
\begin{equation}
    H_{XYZ} = \sum_{i = 1}^{N-1} J_{x} \sigma_{x}^{i}\sigma_{x}^{i + 1} + J_{y} \sigma_{y}^{i}\sigma_{y}^{i + 1} + J_{z} \sigma_{z}^{i}\sigma_{z}^{i + 1},
    \label{eq:Hamiltonian_XYZ}
\end{equation}
where $J_{\alpha}$ stands for the coupling in $\alpha$ direction. Clearly this Hamiltonian does not commute with $S_{z}$ for general choices of $J_{\alpha}$'s. Nonetheless, the XYZ Hamiltonian still supports the mirror symmetry. To perform our VQE analysis, we use the circuit in Fig.~\ref{fig:ansatz}(a) with the difference that $\theta_{x}$, $\theta_{y}$, and $\theta_{z}$ in $\mathcal{N}$ (see Fig.~\ref{fig:circuits}(a)) are replaced by $\theta_{\alpha} = J_{\alpha}\theta$. Note that by this choice, the number of parameters remains the same as the case for the Heisenberg Hamiltonian. 

In Fig.~\ref{fig:XYZ_symmetry_applied}(a) we plot the average energy $\langle H \rangle$ as a function of COI for a circuit of different depths in a chain of length $N=10$ using the strategy 1, namely random initialization of parameters. Clearly, the average energy converges as the number of COI increases and the converged value gets closer to the real ground state energy as the circuit gets deeper. In Fig.~\ref{fig:XYZ_symmetry_applied}(b), we plot the fidelity for a circuit with 6 layers. By increasing the COI, the fidelity increases above the $0.95$ threshold, showing that the protocol still works even in this more general case. To speedup the convergence, in Fig.~\ref{fig:XYZ_symmetry_applied}(c) we plot the average energy $\langle H \rangle$ as a function of COI employing the layer recursive strategy which shows faster convergence in contrast to the random strategy case. To evidence this more clearly, in Fig.~\ref{fig:XYZ_symmetry_applied}(d) we plot the fidelity as a function of COI when the layer recursive strategy is exploited. As evident from the figures, the layer recursive strategy shows significant improvement over the random strategy, which demonstrates that our protocols for speeding up the convergence are still applicable in more general Hamiltonians.

In the XYZ case, as the Hamiltonian supports the mirror symmetry, we can still design the circuit using this constraint, i.e. $\theta_{i} = -\theta_{N - i + 1}$. To go beyond this symmetry, we now consider the spin chain Kondo Hamiltonian~\cite{S_rensen_2007_P08003, S_rensen_2007_L01001, abol2010kondo} as one of the key models in impurity physics. The Hamiltonian can be written as
\begin{equation}
    H_{Kondo} = J\left(J' \boldsymbol{\sigma}^{1}\cdot\boldsymbol{\sigma}^{2} + \sum_{i = 2}^{N-1} \boldsymbol{\sigma}^{i}\cdot\boldsymbol{\sigma}^{i + 1} \right),
\end{equation} 
where the dimensionless parameter $J' < 1$ determines the first qubit as the impurity. The Kondo Hamiltonian commutes with $S_{z}$ but obviously does not have mirror symmetry. In order to simulate the ground state of this Hamiltonian using the VQE algorithm, we again exploit the circuit of Fig.~\ref{fig:ansatz}(a) with the difference that the coupling between the qubits 1 and 2 is different and the local rotations do not follow the mirror symmetry rule, i.e. $\theta_{i} = -\theta_{N - i + 1}$. Due to this, the number of parameters to be optimized increases.

In Fig.~\ref{fig:kondo}(a) we plot the average energy $\langle H \rangle$ as a function of COI for a system length of $N=10$ and different circuit layers using the random initialization strategy. As shown from the figure, increasing the depth of the circuit one can fairly get closer to the ground state energy of the system. To observe this in more detail, in Fig.~\ref{fig:kondo}(b), we plot the fidelity as a function of COI for the case of 4 layers. As evident, in the absence of the mirror symmetry, the ground state can still be simulated with high fidelity. We now attempt to compare the above strategy with the layer recursive one. In Fig.~\ref{fig:kondo}(c), we plot the average energy $\langle H \rangle$ as a function of COI for different circuit layers. As the figure shows, the convergence of the average energy is clearly accelerated for equals circuit depth. Finally, in Fig.~\ref{fig:kondo}(d) we plot the fidelity as a function of COI for the case of 4 layers, showing that the threshold fidelity of $0.95$ is reached faster than the random strategy case.

The results presented above, i.e. simulating the ground state of the XYZ and Kondo Hamiltonian, show that not only our VQE algorithm works well under more general Hamiltonians but also that the layer recursive strategy can still be applied to accelerate the optimization procedure.

\begin{figure}[t]
    \centering
    \includegraphics[width=\columnwidth]{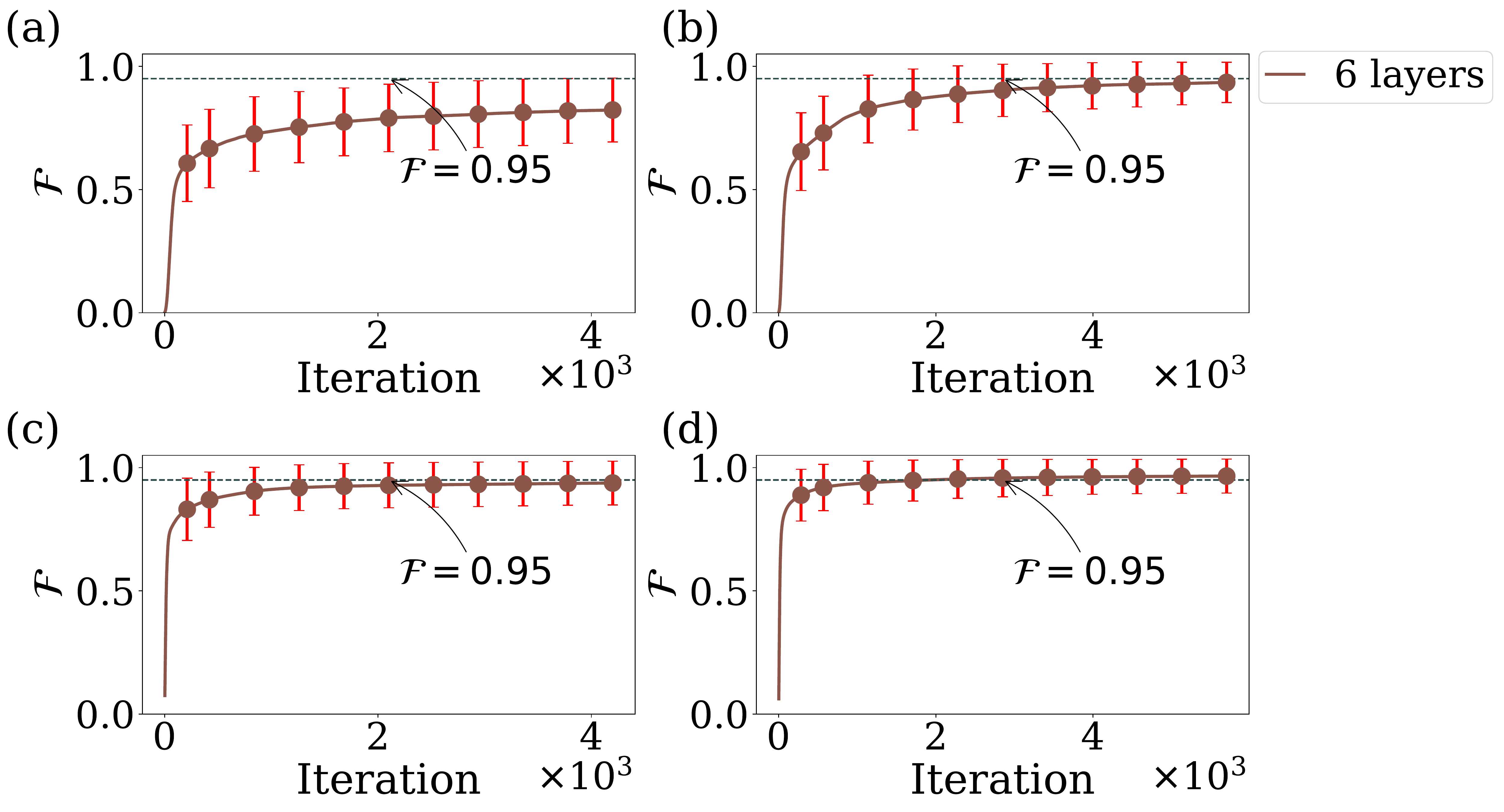}
    \caption{Average obtainable fidelity, and its corresponding error bars, as a function of COI for the XYZ Hamiltonian in a chain of length $N = 10$. (a) Simulating the ground state energy of the system using the random strategy in the presence of mirror symmetry. (b) Same simulation in the absence of the mirror symmetry rule. In panels (c) and (d), we compute the ground state energy using the layer recursive strategy in the presence and in the absence of the mirror symmetry rule, respectively.}
    \label{fig:XYZ_symmetry_not_applied}
\end{figure}

\section{The role of mirror symmetry}

As mentioned before, the presence of mirror symmetry can simplify the circuit through reducing the number of parameters for the local rotations. This has been implemented by the specific choice of rotation angles, i.e. $\theta_{i} = -\theta_{N - i + 1}$. The use of mirror symmetry in designing the circuit has clear advantages as reducing the number of parameters implies the need of less number of measurements for computing the gradient in the optimization procedure. However, one may argue that having a larger parameter space may provide a shorter route to the global minimum. To investigate this, we compare the VQE algorithm for the XYZ Hamiltonian in the presence and in the absence of the mirror symmetry. In Fig.~\ref{fig:XYZ_symmetry_not_applied}(a), we plot the fidelity as a function of COI when the mirror symmetry is employed in our random strategy. While the fidelity improves by increasing the iterations, the final fidelity is not very high. In Fig.~\ref{fig:XYZ_symmetry_not_applied}(b), we plot the same quantity, still using the random strategy, when no mirror symmetry is employed in the circuit and thus the parameter space is larger. Surprisingly, the fidelity shows significant improvement showing that the mirror symmetry may make the convergence slower. To see the performance of our layer recursive strategy, in Fig.~\ref{fig:XYZ_symmetry_not_applied}(c) and (d) we plot the same quantity as a function of iterations in the presence and in the absence of mirror symmetry. Both figures show significant improvement in contrast to the case of random strategy. However, the remarkable observation is that even in the layer recursive strategy still the absence of mirror symmetry in our circuit can reduce the number of iterations. 

It is, however, not easy to conclude that the mirror symmetry is not useful. We have to emphasize that in practical situations for obtaining the gradient of $\langle H \rangle$ with respect to the parameters of the system demands one measurement setup (with thousands of repeatations) for every single parameter. 
Therefore, from an experimental perspective, exploiting the mirror symmetry is beneficial. In addition, if the parameters are too many, the classical optimization may indeed fail. Based on this, although for smaller system sizes, not using the mirror symmetry may demand less iterations, in larger systems it will be more useful to implement such symmetry in the circuit.

\section{Designing the VQE circuit}

The key step for the success of the VQE algorithm is to design a proper circuit such that explores the right part of the Hilbert space which contains the state of interest. In addition, due to practical constraints, e.g. imperfect gates and decoherence, the designed circuit has to be as shallow as possible. Interestingly, the adiabatic circuits can always be considered as a special case of the VQE such that each layer changes the quantum state minimally (as $\Delta t$ is typically small). Motivated by this, one can always adapt the adiabatic circuit, either first or second order Suzuki-Trotter, and add local rotations to all the qubits of each layer, see Fig.~\ref{fig:ansatz}(a). By properly optimizing the local rotations and the parameters of the entangling gates, one can make the operation of each layer significantly different from identity and thus reduce the number of layers. In other words, the adiabatic circuit is an inefficient realization of the VQE, with no local rotations and small action of the entangling gates. Consequently, the operation of each layer in adiabatic circuit is close to identity. Therefore, one can conclude that the VQE always outperforms the adiabatic evolution with respect to the circuit complexity.

\section{Decoherence}

\begin{figure}[t]
    \centering
    \includegraphics[width=\columnwidth]{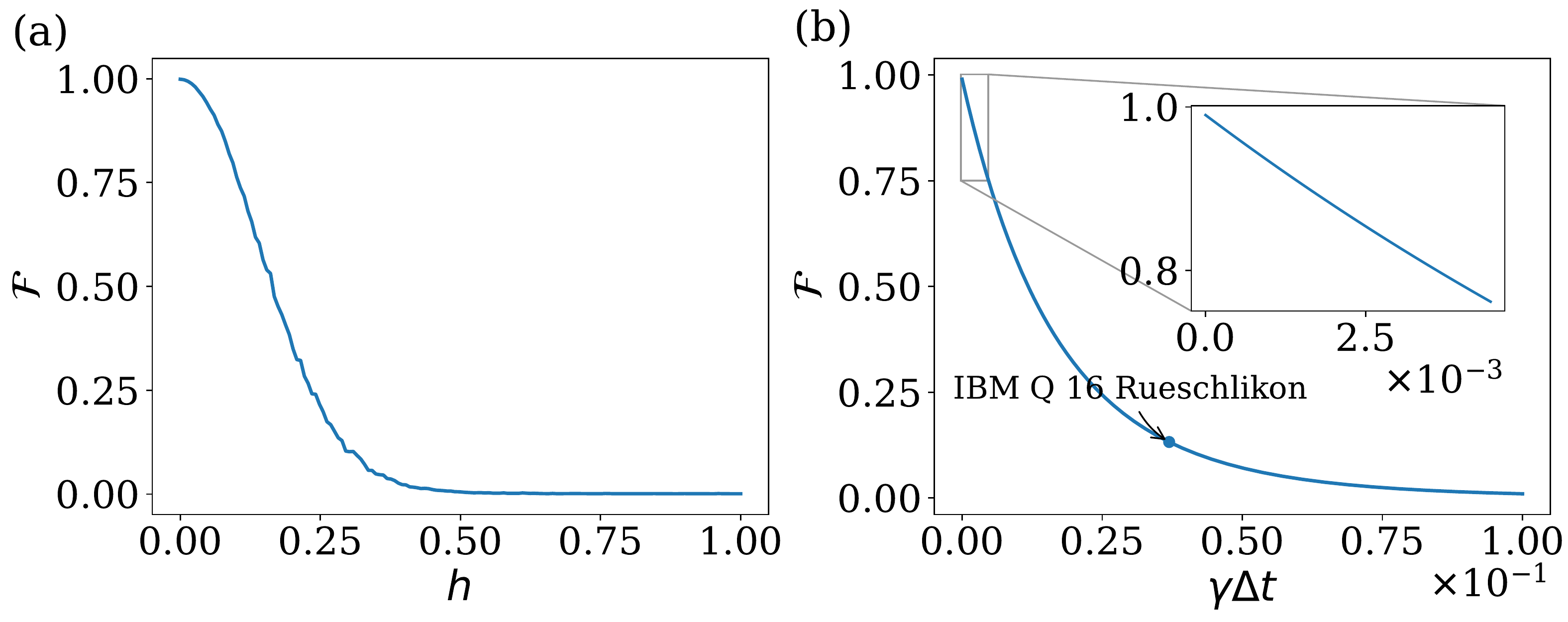}

    \caption{(a) The obtainable VQE fidelity as a function of disorder strength $h$ when the CNOT gates suffer a unitary error, given in Eq.~(\ref{eq:CNOT_Ising}).  (b) The obtainable VQE fidelity as a function of dephasing rate $\gamma \Delta t$. In both cases the system size is $N=10$.}
    \label{fig:decoherence}
\end{figure}

In the NISQ era, the quantum simulators are noisy and their performance is affected by the presence of decoherence. There are several attempts to simulate the effect of noise in existing digital quantum simulators~\cite{li2017efficient,kandala2017hardware,kandala2019error,temme2017error,endo2018practical,maciejewski2020mitigation,premakumar2018error,bonet2018low}. Here, although in a very simplistic and rather naive fashion, we model some errors that can serve as a first glimpse into a more realistic scenario.

We consider two types of errors: (i) unitary ones due to imperfect gate pulses; and (ii) non-unitary errors due to decoherence. Regarding the gate errors, the operations of single qubit rotations are very precise. The performance of CNOT gates, however, may not be as perfect as it expected due to imperfect pulses which couples the two qubits. To see the quality of such errors we consider a particular realization of the CNOT gate through Ising type interaction of the two qubits   
\begin{equation}\label{eq:CNOT_Ising}
\widetilde{U}_{CX}(\phi) = \sqrt{i} \hat{H}_{2} e^{i\sigma^z_{1}\sigma^z_{2}(\pi/4 + \phi)} e^{-i\sigma^z_{1}\pi/4} e^{-i\sigma^z_{2}\pi/4} \hat{H}_{2}
\end{equation}
where $\hat{H}_{2} = (\sigma_2^x+\sigma_2^z)/\sqrt{2}$ is the Hadamard gate, and $\phi$ should ideally be zero but due to imperfect pulses in the device is a random number uniformly distributed over the interval $[-h,+h]$ with zero average. To see the impact of this error, we consider a VQE circuit for a chain of length $N=10$ with $M_{VQE}=3$ layers using optimized parameters, obtained for the ideal case. In reality each CNOT gate will have a random phase $\phi$ in the circuit. We consider $100$ different realization of such random phases and average the final fidelity for each value of $h$, as the noise strength. The average fidelity, using $100$ different random realizations, versus noise strength $h$ is plotted in Fig.~\ref{fig:decoherence}(a). As the figures shows even for a relatively strong noise of $h=0.1$ the fidelity still remains above $0.8$.

A more serious form of noise in NISQ simulators is dephasing which appears as a non-unitary operation and destroys the superposition of quantum states. Dephasing cannot be digitalized and thus it is difficult to incorporate it in our circuit picture. However, in order to have an approximation of its effect, after each layer of gates we evolve the system according to a Lindblad master equation for a certain time  $\Delta t$ which is equal to the time needed to perform the gates in the previous layer. Thus the output of the system after $M$ layers will be
\begin{equation}
\rho^{(M)}=e^{\mathcal{L} \Delta t} \circ U^{(M)} [\rho^{(M-1)}]
\end{equation}
where, $\rho^{(M)}$ ($\rho^{(M-1)}$) represents the density matrix of the system at the output of the layer $M$ ($M-1$), $U^{(M)}$ represents the unitary operation of all the gates in layer $M$ and $\mathcal{L}$ represents the Lindblad master equation $\partial_t\rho=\mathcal{L}(\rho)$ with  
\begin{equation}
\mathcal{L}(\rho)=\gamma \sum_{k=1}^N (\rho-\sigma_k^z \rho \sigma_k^z).
\end{equation} 
The coefficient $\gamma$ shows the strength of the decoherence. In Fig.~\ref{fig:decoherence}(b), we plot the fidelity as a function of $\gamma \Delta t$ for a VQE circuit of length $N=10$ and depth $M_{VQE}=3$ with all the parameters already optimized. As the figure shows up to $\gamma \Delta t =0.0125$  the fidelity remains above $0.5$ which reveals considerable contribution of the ground state in the output of the quantum simulator.

\section{Conclusion}

In this paper, we have investigated two different strategies, namely the adiabatic evolution and the VQE, for simulating the ground state of many-body systems on digital quantum simulators. Our results are multifold. First, for implementing the adiabatic algorithm on a digital quantum simulator, our results show that the second order Suzuki-Trotter circuit demonstrates a clear superiority over the first order by demanding significantly less number of CNOT gates for delivering the same fidelity.
Second, we demonstrate that the VQE approach demands shallower circuits with significantly less number of two-qubit entangling gates in comparison with the adiabatic evolution. This makes the VQE more suitable for the NISQ simulators. However, the simplicity in the VQE circuit comes with the price of the necessity for a classical optimizer which may demand a significant amount of iterations to succeed.
Third, as our main results, we have developed two approaches, namely qubit and layer recursive methods, for accelerating the convergence of the classical optimizer. Both of these approaches, try to start the optimization procedure with an initial guess close to the global minimum and thus reduce the total number of iterations as well as the chance of getting trapped in local minimums. These strategies, indeed, significantly reduce the COI required for the convergence of the VQE algorithms.  

The protocols for accelerating the optimization are very general and work perfectly well for a wide range of Hamiltonians. This includes the Hamiltonians which do not preserve the number of excitations, such as the XYZ, and the ones without the mirror inversion symmetry, such as the single impurity Kondo model. In addition, we have provided a detailed analysis for exploiting symmetries in designing the VQE circuit as well as the destructive role of decoherence. All the codes can be found in \cite{chufan2020}.

\section*{Acknowledgements}
AB acknowledges the National Key R\&D Program of China, Grant No. 2018YFA0306703. VM thanks the Chinese Postdoctoral Science Fund for grant 2018M643435.

\bibliographystyle{abbrvnat}
\bibliography{Adiabatic_VQE}

\end{document}